\documentclass[11pt]{article}
\usepackage{amsmath}
\usepackage{graphicx}
\usepackage{amssymb}

\DeclareMathOperator{\dd}{{\rm d}\!}

\newcommand{\pdf}[2]{\frac{\partial #1}{\partial #2}}
\newcommand{\df}[2]{\frac{\dd #1}{\dd #2}}
\newcommand{\ddx}[1]{\frac{\dd}{\dd {#1}}}
\newcommand{\bsy}[1]{\boldsymbol{#1}}
\title{Small in-plane oscillations of a slack catenary using assumed modes}
\author{Bidhayak Goswami\thanks{Department of Mechanical Engineering, Indian Institute of Technology Kanpur} \and Indrasis Chakraborty\thanks{Center for Applied Scientific Computing (CASC), Lawrence Livermore National Laboratory, US Department of Energy}
 \and Anindya Chatterjee\thanks{Department of Mechanical Engineering, Indian Institute of Technology Kanpur}}
\begin{document}
\maketitle
\begin{abstract}

In this paper we study a problem in oscillations wherein the assumed modes method offers some analytical and theoretical peculiarities. Specifically, we study small in-plane oscillations of a slack catenary, or a sagging inextensible chain fixed at both endpoints. The horizontal and vertical displacements cannot be approximated independently because of pointwise inextensibility in the chain. Moreover, the potential energy is a linear function of the generalized coordinates, and does not directly cause oscillations. Using assumed modes for the vertical displacements only, integrating from one endpoint to compute the required horizontal displacements, treating the horizontal fixity at the distal end as an added scalar constraint, and obtaining linearized equations, we construct an eigenvalue problem which contains a Lagrange multiplier. For generic assumed modes, the Lagrange multiplier is determined by enforcing equilibrium in the undeflected shape. However, when the modes thus determined are reinserted in the assumed mode expansion
and the calculation done afresh, then the Lagrange multiplier is indeterminate at first order. Upon retaining terms at the next order, the distal end fixity constraint introduces quadratic terms into a Lagrangian
without constraints. Results from these two approaches match perfectly. Our approach offers nontrivial insights into both oscillations and Lagrangian mechanics. It is also potentially applicable to other problems with inextensibility in one-dimensional slender members.
  
\end{abstract}

\section{Introduction}

The dynamics of hanging chains has been studied by several authors over more than a century. Routh \cite{routh1905advanced} wrote equations for the uniform chain but provided analytical solutions for a non-uniform chain whose equilibrium shape is a cycloid. The equilibrium shape of a sagging inextensible chain with uniform mass per unit length is in fact a hyperbolic cosine (see, e.g., \cite{gelfand1963calculus}). 
Pugsley \cite{pugsley1949natural} conducted experiments and gave semi-empirical expressions for first three natural frequencies of a catenary. Saxon and Cahn \cite{saxon1953modes} studied the problem analytically and gave an asymptotic solution for small sag to span ratio. Their results were in agreement with Pugsley \cite{pugsley1949natural}. An approximate solution of equations from Routh \cite{routh1905advanced} and Saxon et al.\ \cite{saxon1953modes} was found by Goodey \cite{goodey1961natural}. The expressions obtained were close to the empirical formulas of Pugsley \cite{pugsley1949natural}. 

The catenary with small sag to span ratio leads to an apparent contradiction, as noted by Irvine and Caughey \cite{irvine1974linear}.
For very small sag to span ratios, the first mode is antisymmetric. However, for a flat and taut string, the first mode is symmetric. The contradiction arises because at extremely small sag, the tension is very high, and an extensible-chain treatment is needed to obtain the flat-string behavior. However, in this paper, we are not interested in that regime: we examine the catenary in the regime where the sag to span ratio is neither very small nor very large.
 
With newer computational techniques, more detailed studies of more complex chain systems were possible. Simpson \cite{simpson1966determination} used a transfer matrix approach to study a multi-span transmission line.
Finite Element Analysis (FEA) was used as well \cite{gambhir1979finite,ahmadi1989vibration}. Karoumi \cite{karoumi1999some} examined a model with catenary cables supporting a bridge deck.

Rega \cite{rega2004nonlinear,rega2004nonlinear2} has presented a detailed discussion of the dynamics of {\em extensible} cables beginning from a continuum mechanics framework and considering both various analytical simplifications as well as detailed solution aspects only obtainable using numerical methods. The reader may refer to these excellent review papers for many more references relevant to the dynamics of a catenary and related problems.

In contrast to the above papers which either allow extensibility, or assume small sag, or discretize the system using finite elements and move to a fully numerical treatment, in this paper we address the strictly inextensible catenary, with large sag, using the Lagrangian approach and assumed
modes\footnote{%
		The assumed modes approach is also called the Rayleigh-Ritz approach.}.
The assumed modes approach has enjoyed wide adoption in mechanics because of its conceptual simplicity, ease of refinement to useful accuracy, focus on essential kinematics, sidestepping of natural boundary conditions, and direct use in the Lagrangian approach. The interested reader may see a few examples in the following papers: drum vibrations in
\cite{bridge2007vibration}, rotor dynamics in \cite{chun1996vibration,lee1998vibration}, flexible multibody systems in \cite{tadikonda1995assumed}, an application in piezoelectric energy harvesting in \cite{erturk2012assumed}, the dynamics of flexible robots in \cite{celentano2011computationally}, among many more in the literature.
We note that the catenary itself continues to appear in research papers devoted to more complicated interactions, such as cable robots \cite{d2021catenary} and catenary risers in marine applications \cite{zhu2021spatial}.

Following the above discussion, we can now introduce the problem studied in this paper and point out its interesting features to motivate our work. We study small in-plane oscillations of an inextensible catenary with considerable sag, i.e., a slack catenary. We begin with the well known static
equilibrium configuration of the catenary. We use the horizontal spatial coordinate as an independent variable during the initial setup of the problem.
After introducing assumed modes for small vertical displacements of the chain, as is usual within the Lagrangian formulation, we retain only time as an independent variable. The pointwise inextensibility of the chain
relates small vertical and horizontal displacements through a differential equation, which we can integrate to obtain the horizontal displacements. Taking one endpoint of the chain to be fully fixed and the vertical location of the other, or distal, endpoint to be fixed as well, it is convenient to impose the horizontal location of the distal endpoint
through an explicit scalar constraint equation which introduces a Lagrange multiplier. A key aspect of this approach is that the potential energy of the system is {\em linear} and not {\em quadratic} in the displacements, and so the equations of motion obtained using the Lagrangian approach have some nonzero terms, including the Lagrange multiplier, that survive even when the displacements are set to zero. Since zero displacement represents equilibrium, these nonzero terms must necessarily add up to zero. Setting their sum to zero determines the Lagrange multiplier. Physically, this means the frequency remains indeterminate until horizontal fixity of the second endpoint of the chain is enforced. Furthermore, in a subsequent calculation,
when we do not use arbitrary shape functions in the assumed modes and instead use the correct modes as determined in the first part, then those erstwhile nonzero terms become identically zero and the Lagrange multiplier becomes indeterminate within the linear approximation. In this case, retaining quadratic terms in a nonlinear treatment leads to a potential energy which includes quadratic terms, and the problem becomes determinate. In this way, this seemingly simple classical problem leads to clear, potentially useful, and in our opinion pleasing academic insights into both
oscillations as well as the role of constraints within Lagrangian mechanics.

\section{Problem setup}\label{the_catenary}
\begin{figure}[h!]
\centering
\includegraphics[scale=0.4]{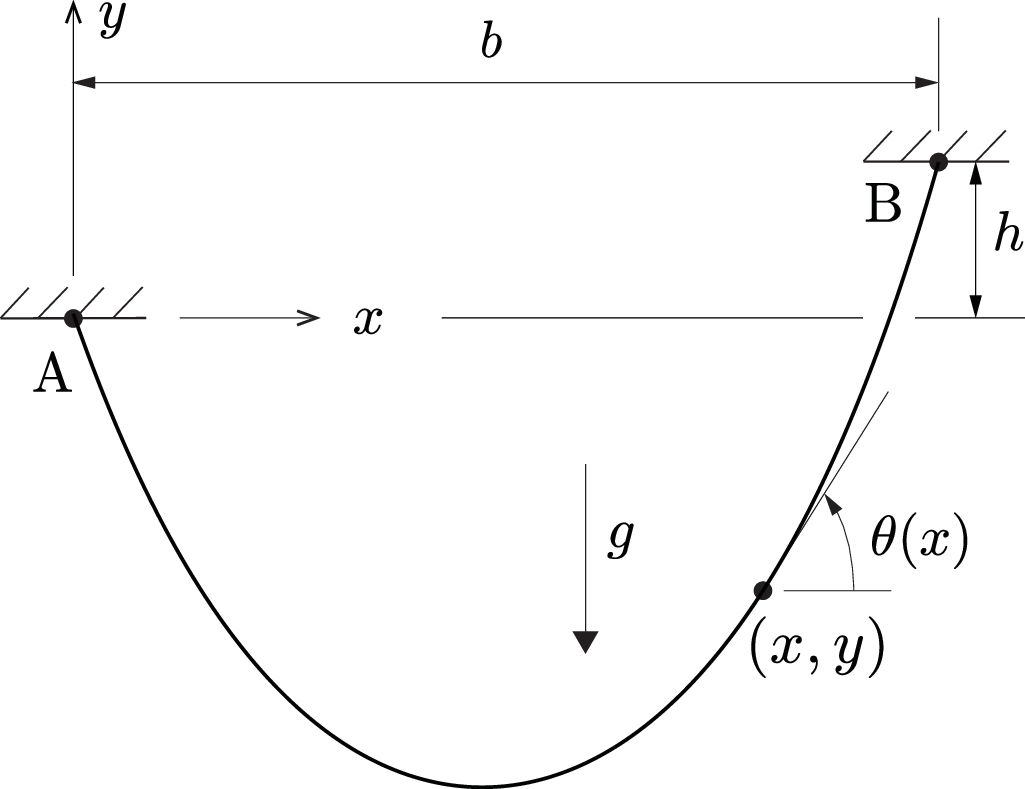}
\caption{A slack catenary hanging from two fixed points. Gravity acts downward. Without loss of generality, we take $h \ge 0$.}
\label{fig_1_catenary}
\end{figure}
We consider an inextensible, slack catenary (Fig.\ \ref{fig_1_catenary}) of length \(L\) and a uniform mass \(\bar{m}\) per unit length, hanging from two fixed ends A and B. We use a coordinate system with origin at A. The end B is at the point \((b,h)\), with $h \ge 0$. 
Gravity acts along the negative \(y\) direction. The tangent to the equilibrium curve at location \(x\) makes an angle \(\theta(x)\) with the positive \(x\) direction. Denoting the equilibrium shape as $y(x)$, we have
\begin{equation}
	y(0)=0\quad \mbox{ and }\quad y(b)=h.
\end{equation}
By nondimensionalization or by choice of units, we take \(L=1\), \(\bar{m}=1\), and $g=1$. Although our treatment is general, analytical intractability
forces us to use numerical integrals. So we will study two cases in detail: $h=0$ and $h=0.1$, with $b=0.6$ in both cases.

\subsection{Equilibrium}\label{eq_sol}
If \(T\) is the spatially varying tension in the chain, equilibrium requires
\begin{equation}\label{tcos}
	\ddx{x}\left(T\cos(\theta)\right)=0,\mbox{ and }
\end{equation}
\begin{equation}\label{tsin}
	\ddx{x}\left(T\sin(\theta)\right)=\bar m\,g\,\sec(\theta).
\end{equation}
From Eq.\ \ref{tcos}, \(T\cos(\theta)=T_0\), a constant, whence Eq.\ \ref{tsin} yields
\begin{equation}\label{eql_ode}
	\frac{{\rm d}^2 y}{\dd x^2}=\frac{1}{W}\,\sqrt{1+\left(\df{y}{x}\right)^2}=\frac{1}{W}\,\sec(\theta),\quad \mbox{where }W=\frac{T_0}{\bar m\,g}.
\end{equation}
Solution of Eq.\ \ref{eql_ode} with $y(0)=0$ yields
\begin{equation}
	y=-W\cosh(C)+W \cosh\left(C+\frac{x}{W}\right),
\end{equation}
where \(W\) and \(C\) are constants to be determined numerically from the boundary condition $y(b)=h$ and the length condition
\begin{equation}
	\int_{0}^{b}\sec(\theta)\dd x=\int_{0}^{b}\sqrt{1+\left(\frac{\dd y}{\dd x}\right)^2}\dd x=L.
\end{equation}

For example, with $b=0.6$ and $h=0$, we have $W=0.1631683$ and $C=-1.8385927$; and with $b=0.6$ and $h=0.1$, we have \(W=0.1640525\) and \(C=-1.7283471\).

\subsection{Inextensibility constraint}
\label{inextensibility}

Let \(u(x,t)\) and \(v(x,t)\) be the displacement components along \(x\) and \(y\) directions respectively. 
The length of a small element, even after displacement, remains unchanged.
\begin{figure}[h!]
	\centering
	\includegraphics[scale=0.4]{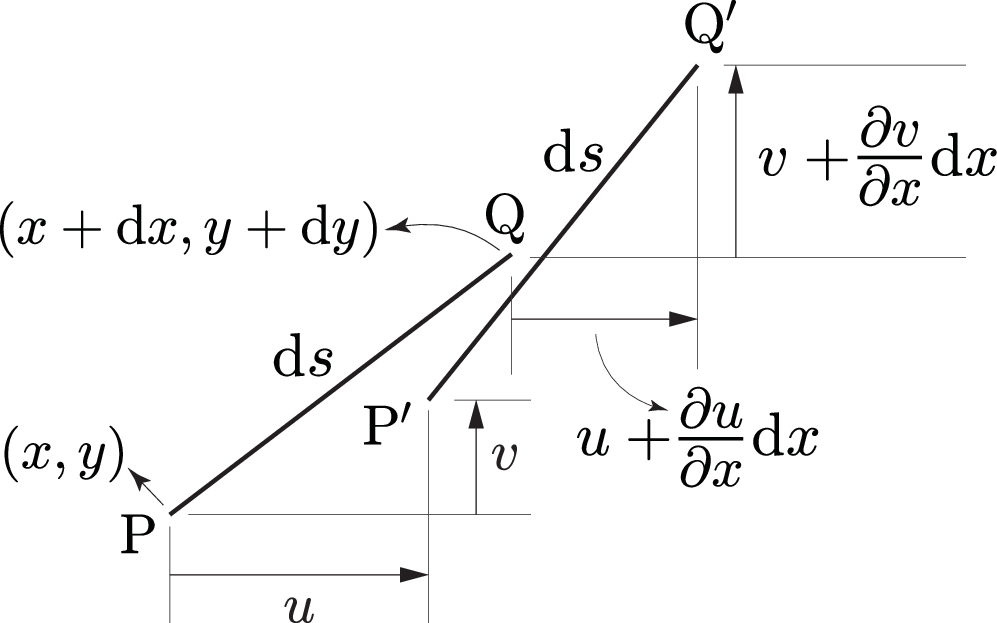}
	\caption{An element of length `\(\dd s\)' in equilibrium and displaced configuration.}
	\label{inex}
\end{figure}
In Fig.\ \ref{inex}, the endpoints P and Q of an element of length \(\dd s\) move to \({\rm P}'\) and \({\rm Q}'\). Inextensibility requires
\begin{equation}\label{inexeq1}
	{\dd s}^2={\dd x}^2+{\dd y}^2 = \left(1+\pdf{u}{x}\right)^2{\dd x}^2+\left(\df{y}{x}+  \pdf{v}{x}  \right)^2{\dd x}^2.
\end{equation}
Using subscripts and primes to denote partial and ordinary derivatives with respect to $x$, we obtain the pointwise {\em differential} constraint
\begin{equation}\label{inex_u}
	1+{y'}^2=\left(1+u_x\right)^2+\left(y'+v_x\right)^2
\end{equation}
which simplifies to
\begin{equation}\label{inex_ux_original}
	{u_x}=\sqrt{1-2\,y'\,{v_x}-{v_x}^2}-1,
\end{equation}
where the sign on the square root is chosen to ensure that $u$ is infinitesimal when $v$ is infinitesimal.
Note that, if $v$ is expanded using assumed modes, then $u$ can be found by integration with respect to $x$ so that inextensibility is obeyed pointwise.

\subsection{Horizontal fixity of the right endpoint}
\label{hc}
In an assumed modes approach, we can easily enforce $v(0,t)=v(b,t)=0$ by using, e.g., a truncated Fourier sine series. Substituting that expression
into Eq.\ \ref{inex_ux_original}, we can in principle integrate with respect to $x$ to obtain $u(x,t)$. Enforcing $u(0,t)=0$ is done by taking the lower limit of the integral to be $x=0$. However, for a general assumed modes expansion, $u(b,t)=0$ is not guaranteed and must be enforced as an additional constraint.

Since we are interested in small oscillations of the chain, we assume \(u,\,v\), and their spatial derivatives to be small. 
For small \(v_x\), Eq.\ \ref{inex_ux_original} has a Taylor series approximation, i.e.,
\begin{equation}\label{inex_ux}
	u_x=-{\it y'}\,{v_x}-\frac{1}{2}\left(1+{y'}^2 
	\right) {{v_x}^2}+\cdots .
\end{equation}
Integrating Eq.\ \ref{inex_ux} in space, the horizontal component \(u\) can be approximated using
\begin{equation}\label{u_approx}
	u=\int_{0}^{x}\left( -{\ y'(\xi)}\,{v_\xi}-\frac{1}{2}\left(1+{y'(\xi)}^2 
	\right) {{v_\xi}^2}+\cdots    \right)\dd\xi.
\end{equation}
We now introduce our assumed modes
\begin{equation}\label{v_assumed_mode}
	v(x,t)=\sum_{k=1}^{N}a_k(t)\sin\left(\frac{k\pi x}{b}\right),
\end{equation}
as mentioned above. The corresponding horizontal displacement \(u\) does not automatically satisfy $u(b,t)=0$. We therefore include the constraint
\begin{equation}\label{holonomic}
	u(b,t)=\int_{0}^{b}\left( -{\ y'(x)}\,{v_x}-\frac{1}{2}\left(1+{y'(x)}^2 
	\right) {{v_x}^2}+\cdots   \right)\dd x=0.
\end{equation}
Although the constraint is holonomic, we do not use it to eliminate a degree of freedom because we do not wish to make {\em a priori} assumptions
about which term in the Fourier sine series can be eliminated. We now proceed to write equations of motion.

\section{Equations of motion}\label{equ_motion}

The kinetic energy of the catenary  is 
\begin{equation}\label{KE}
	\mathcal{T}=\frac{1}{2}\int_{0}^{b}\bar{m}\left(\dot{u}^2+\dot{v}^2\right)\sec(\theta)\dd x
\end{equation}
where the overdot denotes a partial derivative with respect to time, and the potential energy of the catenary, after subtracting a constant corresponding to the equilibrium configuration, is
\begin{equation}\label{PE}
	\mathcal{V}=\int_{0}^{b}\bar{m}\, g\, v\, \sec(\theta) \dd x.
\end{equation}
The Lagrangian 
\begin{equation}
	\mathcal{L}=\mathcal{T}-\mathcal{V},
\end{equation}
and the equations of motion are 
\begin{equation}\label{eqm}
	\ddx{t}\left(\pdf{\mathcal{L}}{\dot{a}_k}\right)-\pdf{\mathcal{L}}{a_k}=\lambda\, \frac{\partial}{\partial a_k}\left(u(b,t)\right), \quad k=1,2,\dots,N.
\end{equation}
In the above, $u(b,t)$ stands for the definite integral in Eq.\ \ref{holonomic}, and
\(\lambda\) is to be determined as part of the solution. For small oscillations, we want equations that are correct up to first order in the generalized coordinates $a_k$. This means that for computing the kinetic energy in Eq.\ \ref{KE}, the integral of Eq.\ \ref{u_approx} must be evaluated
only up to first order in the $a_k$. However, because the expression for $u(b,t)$ in the constraint equation is differentiated once, the integral of Eq.\ \ref{holonomic} must be evaluated up to second order in the $a_k$. The latter takes the form
\begin{equation}
\label{ceq}
	u(b,t)=\bsy{a}^\top \bsy{q}+\frac{1}{2}\,\bsy{a}^\top\, {\bf B}\,\bsy{a}+ \cdots = 0,
\end{equation}
where  \(\bsy{a}=[a_1, a_2,\dots,a_N]^\top\) and ${\bf B}$ is a constant symmetric matrix of size \(N\times N\), or
${\bf B}\in{\mathbb{R}}^{N \times N}$.
Then Eqs.\ \ref{eqm} take the form
\begin{equation}\label{eqm1}
	{\bf M}\,\ddot{\boldsymbol{a}}+\bsy{p}=\lambda\,\bsy{q}+\lambda\, {\bf B}\,\bsy{a},
\end{equation}
where the constant vectors
\(\bsy{p}\) and \(\bsy{q}\) are generally nonzero (the degenerate case will be discussed in Section \ref{exact_modal_expansion}). Since $\bsy{a}$ is zero at equilibrium, the terms in \(\bsy{p}\) and \(\bsy{q}\) must cancel exactly at equilibrium. We therefore expect that \(\bsy{p}\) and \(\bsy{q}\) are parallel, as indeed they are (see appendix \ref{app2_pq_parallel}).
In Eq.\ \ref{eqm1},  \({\bf M}\) is a symmetric positive definite matrix of size \(N\times N\).

We must now determine $\lambda$ correct up to linear terms. Considering equilibrium in Eq.\ \ref{eqm1}, let  \(\bsy{a}=\bsy{0}\), and
 \(\lambda=\lambda_0\). Then
\begin{equation}\label{pq_rel}
	\bsy{p}=\lambda_0\,\bsy{q}.
\end{equation}
If the vectors \(\bsy{p}\) and \(\bsy{q}\) are nonzero, their parallelism ensures that a unique \(\lambda_0\) can be determined. Having determined \(\lambda_0\), we can consider small motions. For small \({\bsy{a}}\), let
\begin{equation}\label{lmbda_val}
	\lambda=\lambda_0+\eta
\end{equation}
where \(\eta\) is of the same order of magnitude as \(\lvert\lvert{\bsy{a}}\rvert\rvert\). Substituting Eq.\ \ref{lmbda_val} in Eq.\ \ref{eqm1}, and using Eq.\ \ref{pq_rel}, we obtain up to first order,
\begin{equation}\label{eqm2}
	{\bf M}\,\ddot{{\bsy{a}}}=\eta\, \bsy{q}+\lambda_0\,{\bf B}\,{\bsy{a}}.
\end{equation}
Linearizing Eq.\ \ref{ceq}, we see that the vector \(\bsy{a}\) must lie on \(\mathcal{X}\), the \((N-1)\) dimensional subspace of \(\mathbb{R}^N\) that is orthogonal to \(\bsy{q}\). Hence, we write 
\begin{equation}\label{asub}
	\bsy{a}={\bf Q}\,\bsy{\zeta},
\end{equation}
where matrix \({\bf Q}\), of size $N \times (N-1)$,  provides a basis for \(\mathcal{X}\). Note that in the software package Matlab, such a
\({\bf Q}\) can be easily obtained using the ``qr'' decomposition. We now have 
\begin{equation}
	{\bf Q}^\top\bsy{q}=\bsy{0}.
\end{equation} 
The vector \(\bsy{\zeta}=[\zeta_1,\zeta_2,\dots,\zeta_{N-1}]^\top\) contains time varying coordinates in the new basis. Substituting Eq.\ \ref{asub}  in Eq.\ \ref{eqm2}, and premultiplying with \({\bf Q}^\top\), we obtain
\begin{equation}\label{reduced_evp}
	\tilde{\bf M}\, \ddot{\boldsymbol{\zeta}}=\lambda_0\,\tilde{\bf B}\,\bsy{\zeta},
\end{equation}
where \(\tilde{\bf M}={\bf Q}^\top\,{\bf M}\,{\bf Q}\) and \(\tilde{\bf B}={\bf Q}^\top\,{\bf B}\,{\bf Q}\), and
$\tilde{\bf M},\tilde{\bf B}\in{\mathbb{R}}^{(N-1)\times(N-1)}.$
The approximated frequencies and mode shapes of the catenary can now be calculated by solving the eigenvalue problem in Eq.\ \ref{reduced_evp}.

\section{Numerical results}
\label{numres}
We now present numerical results for two cases mentioned in Section \ref{the_catenary}. Numerical calculations were carried out to several digits more than those displayed.
\subsection{Case 1: $h=0$}
\label{symcase}
This case refers to a catenary which is symmetrically suspended between two points. We obtain
\begin{equation}
	{\bf M}=\begin{bmatrix}
		3.4764 &   2.5342 &   1.6515 &   2.2369\\
		2.5342 &   3.3974 &   1.7375 &   2.4056\\
		1.6515 &   1.7375 &   2.3715&    1.5337\\
		2.2369  &  2.4056  &  1.5337 &   3.1275
	\end{bmatrix},\,
	{\bf B}=\begin{bmatrix}
		-38.9759   &      0 & -55.2220  &       0\\
		0 &-127.3652  &       0& -125.5176\\
		-55.2220   &      0& -262.1130    &     0\\
		0 &-125.5176 &       0& -447.1589
	\end{bmatrix},
\end{equation}
and 
\begin{equation}
	\bsy{p}=\begin{Bmatrix}
		0.5195\\
		0\\
		0.3562\\
		0
	\end{Bmatrix},\,
	\bsy{q}=\begin{Bmatrix}
		3.1838\\
		0\\
		2.1830\\
		0
	\end{Bmatrix}.
\end{equation}
From Eq.\ \ref{pq_rel}, $ \lambda_0=0.1631682 $. The first three angular frequencies are found to be  2.4294,   4.3590, and   6.1950. 
\begin{figure}[h!]
	\centering
	\includegraphics[width=\linewidth]{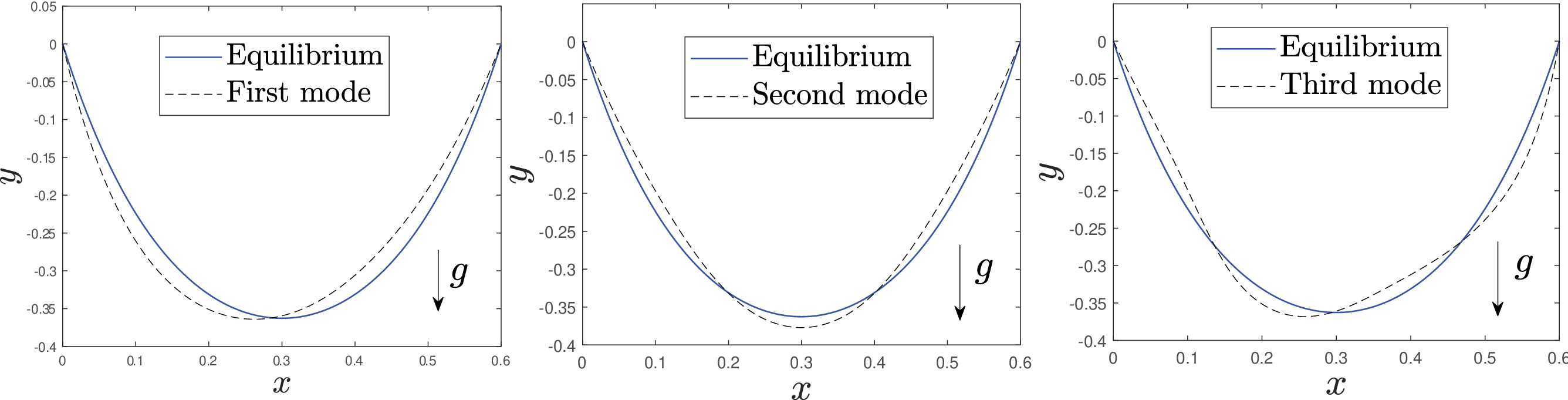}
	\caption{First three mode shapes of the in-plane oscillations of the catenary for $h=0$. The solid lines show the equilibrium shape.}
	\label{sy_mode}
\end{figure}
The corresponding mode shapes are shown in Fig.\ \ref{sy_mode}.

\subsection{Case 2: $ h=0.1 $}
This case refers to a catenary which is asymmetrically suspended between two points. We obtain
\begin{equation}
	{\bf M}=\begin{bmatrix}
		3.2559 &   2.1171&    1.2612 &   2.0685\\
		2.1171 &   2.9911  &  1.1606  &  2.0895\\
		1.2612 &   1.1606  &  1.9750 &   0.9706\\
		2.0685 &   2.0895  &  0.9706 &   2.9150
	\end{bmatrix},\,
	{\bf B}=\begin{bmatrix}
		-39.1116 &  10.7499 & -55.2174 &   9.7637\\
		10.7499& -127.8014 &  27.0240& -125.5289\\
		-55.2174 &  27.0240& -263.1210&   49.8938\\
		9.7637& -125.5289&   49.8938& -449.0069
	\end{bmatrix},
\end{equation}
and
\begin{equation}
	\bsy{p}=\begin{Bmatrix}
		0.5205\\
		-0.0435\\
		0.3551\\
		-0.0268
	\end{Bmatrix}, \,\,
	\bsy{q}=\begin{Bmatrix}
		3.1725\\
		-0.2650\\
		2.1648\\
		-0.1636
	\end{Bmatrix}.
\end{equation}
From Eq.\ \ref{pq_rel}, $ \lambda_0=0.1640525 $. The first three angular frequencies are 2.4375,   4.3952, and   6.2196. 
\begin{figure}[h!]
	\centering
	\includegraphics[width=\linewidth]{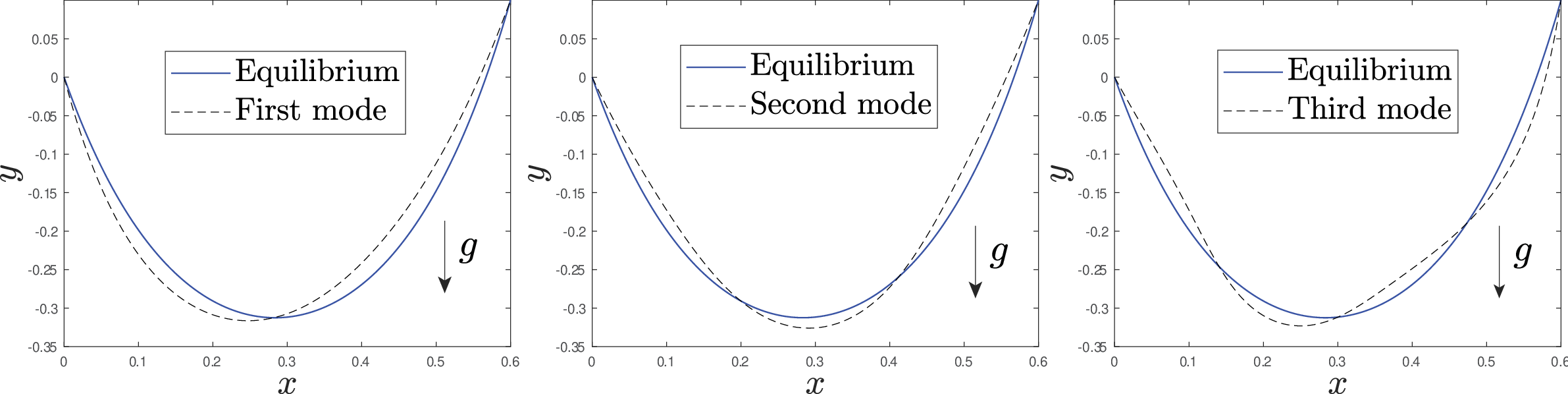}
	\caption{First three modes of the asymmetrically suspended catenary with $h=0.1$. The solid lines show the equilibrium shape.}
	\label{asy_modes}
\end{figure}
The corresponding mode shapes are shown in Fig.\ \ref{asy_modes}.

\section{Degeneracy encountered with actual modes}\label{exact_modal_expansion}
In usual applications of the Rayleigh Ritz approach with assumed modes, exact results are obtained when the exact mode shape is used. In this problem, however, use of the mode shapes obtained above leads to an interesting degenerate condition. Specifically, the constraint Eq.\ \ref{ceq} is satisfied to first order, which means $\bsy{q}$ is zero. There is no inconsistency, and $\bsy{p}$ is zero as well. However, now Eq.\ \ref{pq_rel} cannot be solved for $\lambda_0$.
In other words, when the very modes obtained above are used in the assumed modes calculation in the form
\begin{equation}\label{v_modal}
	v=a_1\,\psi_1 + a_2\,\psi_2 + a_3\,\psi_3 + \cdots,
\end{equation}
then we obtain
\begin{equation}\label{eqm_exact}
	{\bf M}\,\ddot{{\bsy{a}}}=\lambda\, {\bf B}\,\bsy{a},
\end{equation}
with the vectors \(\bsy{p}\) and \(\bsy{q}\) becoming individually zero. Now $\lambda$ is indeterminate within the linear formulation.

The reason for this loss of the $\lambda$-determining vectors \(\bsy{p}\) and \(\bsy{q}\) is not that mode shapes are used. In fact, any shape $v$ which satisfies the fixity condition of both endpoints, i.e., which satisfies the constraint Eq.\ \ref{holonomic}, will lead to this situation. The reason is that these displacements are measured from the equilibrium position; and the equilibrium position is at minimum potential energy. Thus, {\em any} displacement that satisfies the essential boundary conditions (fixities at both ends) must necessarily show no change in potential energy at first order. Near the equilibrium position, every displacement must be accompanied by a locally quadratic rise in potential energy.

Thus, if we use assumed modes that satisfy Eq.\ \ref{ceq}, e.g., if we use the approximate mode shapes determined above as the assumed modes to begin with, then a nonlinear treatment is needed. We now develop the same. A full expansion is tedious, with a large number of terms to handle, so we demonstrate the idea by doing a nonlinear treatment of an expansion close to one of these already-determined modes from Sections \ref{equ_motion} and \ref{numres}.

\section{Nonlinear treatment using one mode}\label{perturb}

In the previous sections we have calculated the mode shapes (to a very good approximation), and have observed that an expansion using the modes themselves, as in Eq.\ \ref{v_modal}, satisfies the constraint (Eq.\ \ref{holonomic}) up to first order of the \(a\)'s. 
We now consider a \(v\) which is primarily along a single mode, but with a small correction term that is needed to satisfy the boundary condition
upto second order in an appropriate expansion. Using $\varepsilon$ as a bookkeeping parameter to keep track of sizes, we write
\begin{eqnarray}\label{v_np}
	v=\varepsilon\, \alpha(t)\,\psi(x)+\varepsilon^2\,\bar{v}(x,t)
\end{eqnarray}
where \(\psi(x)\) is a mode (as obtained above: technically an approximation, but a good one), \(\alpha(t)\) is the primary time-varying coordinate associated with \(\psi(x)\), and the \({\bar v}(x,t)\) a correction that satisfies the zero boundary conditions of vertical displacement at both endpoints\footnote{%
	Note that the horizontal displacements induced by this choice of $v$ will be required to vanish at both endpoints also.}.
If we calculate the potential energy \(\tilde{\cal V}\) from Eq.\ \ref{PE}, there is no direct contribution from the $\psi(x)$ part, as explained above. Hence, the potential energy depends on the second order correction term,
\begin{equation}\label{PE2}
	\tilde{\cal V}=m\,g\,\varepsilon^2\int_{0}^{b}\bar{v}\,\sec(\theta(x)) \dd x.
\end{equation} 

We now turn to the horizontal displacement constraint, Eq.\ \ref{holonomic}, which after integration by parts yields
\begin{equation}\label{holonomic_2}
	u(b,t)=\int_{0}^{b}\left( {\ y''(x)}\,v-\frac{1}{2}\left(1+{y'(x)}^2 
	\right) {{v_x}^2}+\cdots   \right)\dd x .
\end{equation}
In the above, from Eq.\ \ref{eql_ode},
$$y''(x) = \frac{1}{W} \,\sec(\theta(x)),$$
and so (as explained using the potential energy calculation),
$$\int_{0}^{b} y''(x) \,\psi(x) \, \dd x = 0.$$
Therefore, correct up to second order, we have
\begin{equation}\label{holonomic_3}
	u(b,t)=\varepsilon^2\int_{0}^{b}\left( \frac{1}{W}\,\sec(\theta(x))\,\bar{v}-\frac{1}{2}\left(1+{y'(x)}^2 
	\right)\alpha^2{\psi'(x)}^2\right)\dd x =0.
\end{equation}
The first term in the integrand above is directly related to the integrand in the expression for the potential energy in Eq.\ \ref{PE2}, and 
we directly obtain
\begin{equation}\label{PE3}
	\tilde{\cal V}=\frac{W\,m\,g\,\varepsilon^2}{2}\int_{0}^{b}\left(1+{y'(x)}^2 
	\right)\alpha^2{\psi'(x)}^2 \dd x={\cal C}_1\varepsilon^2\alpha^2.
\end{equation}
For the kinetic energy to be evaluated correct up to second order, $\bar v$ can be dropped from Eq.\ \ref{v_np}. 
So the function $\bar v$ need not be determined at this order, after all, and we have
\begin{equation}\label{KE2}
\tilde{\cal T}=\frac{1}{2}\int_{0}^{b}\bar{m}\left(\dot{u}^2+\dot{v}^2\right)\sec(\theta(x))\dd x={\cal C}_2\,{\varepsilon}^{2}\,{{\dot \alpha}}^{2}.
\end{equation}
Now the Lagrangian 
\begin{equation}
	\tilde{\cal L}=\tilde{\cal T}-\tilde{\cal V}={\cal C}_2\,\varepsilon^2\,\dot{\alpha}^2-{\cal C}_1\,\varepsilon^2\,\alpha^2,
\end{equation}
yielding
\begin{equation}\label{har_os}
	\ddot{\alpha}+{\tilde \omega}_1^2\,\alpha=0,
\end{equation}
where
$${\tilde{\omega}_1}=\sqrt{\frac{{\cal C}_1}{{\cal C}_2}},$$
which upon evaluating the integrals turns out to be 2.4294 for the symmetric case, matching perfectly (recall section \ref{symcase}).

\section{Conclusion}

A catenary, or an inextensible chain suspended from two endpoints, has the shape of a hyperbolic cosine at equilibrium. Due to pointwise inextensibility, the vertical and horizontal displacement components are related by a differential equation. An assumed mode solution for small oscillations of a slack catenary presents some challenges and has been missing from the literature. Here, starting with an assumed mode expansion for the vertical displacement, the horizontal displacement was obtained using the inextensibility condition. The horizontal fixity at the distal end
was enforced using an additional scalar constraint. Subsequently, a Lagrangian formulation was used. One of the interesting aspects of this problem is that, in the Lagrangian formulation, the potential energy is {\em linear} in the generalized coordinates. Enforcement of the constraint through a Lagrange multiplier makes the oscillation frequencies determinate.
Further, when these same modes, or any other assumed modes that satisfy the fixity constraint up to first order, are used to expand the vertical displacement, then degeneracy is encountered at first order. However, adding a small perturbation to that assumed mode, and carrying out the calculation to second order, gives the usual harmonic oscillator equation and a fully satisfactory solution.

The method is semi-numerical and can be implemented in software like Matlab and Maple. The treatment here helps to provide interesting insights that may carry over to some other problems, such as the vibrations of a pre-bent elastica.

\section{Conflict of interest}

There are no conflicts of interest.

\appendix

\section{Parallelism of vectors \(\bsy{p}\) and \(\bsy{q}\)}\label{app2_pq_parallel}

The leading order term of Eq.\ \ref{holonomic} is 
\begin{align}
	\int_{0}^{b}-y'(x)\,v_x \dd x&=-y'(x)\,v\,\bigg\rvert_0^b+\int_{0}^b y''(x)\,v \dd \xi\nonumber\\
	&=\int_{0}^b y''(x)\,v \dd x\nonumber\\
	&=\int_{0}^b \frac{1}{W} \,\sec(\theta(x)) \, v\dd x\quad \mbox{(referring Eq.\ \ref{eql_ode})}\nonumber\\
	&=\frac{1}{W} \int_{0}^b \sec(\theta(x))\,v\dd x.
\end{align}
The potential energy is 
\begin{equation}
	{\cal V}=\int_{0}^b \bar{m}\,g\,v\,\sec(\theta(x))\dd \xi=\bar{m}\,g\int_{0}^b \sec(\theta(x))\,v\dd \xi.
\end{equation}
Since the same $v$ appears in both integrals above, they differ only up to a multiplicative constant. This is why \(\bsy{p}\) and \(\bsy{q}\)
are parallel.

\end{document}